\documentclass[conference]{IEEEtran}
\IEEEoverridecommandlockouts

\usepackage{subcaption}
\usepackage{amsmath,amssymb,amsfonts}
\usepackage{graphicx}
\usepackage{textcomp}
\usepackage{xcolor}
\usepackage{flushend}
\usepackage[T1]{fontenc}
\usepackage{dsfont}
\usepackage{balance}
\usepackage{tabulary}
\usepackage{booktabs}
\usepackage{graphicx}
\usepackage{todonotes}
\usepackage[linesnumbered,ruled,vlined]{algorithm2e}
\usepackage{tikz}
\usepackage{xtab}
\usepackage{hyperref}
\newcolumntype{P}[1]{>{\RaggedRight}p{#1}}
\usepackage{soul}
\usepackage{listings}

\def\BibTeX{{\rm B\kern-.05em{\sc i\kern-.025em b}\kern-.08em
    T\kern-.1667em\lower.7ex\hbox{E}\kern-.125emX}}

\makeatletter
\newcommand{\newlineauthors}{%
\end{@IEEEauthorhalign}\hfill\mbox{}\par
\mbox{}\hfill\begin{@IEEEauthorhalign}
}
\makeatother

\newcommand*\circled[1]{\tikz[baseline=(char.base)]{
            \node[shape=circle,draw,inner sep=1pt] (char) {#1};}}

\definecolor{LightGray}{gray}{0.9}
\lstdefinestyle{mystyle}{
	backgroundcolor=\color{LightGray},
	basicstyle=\footnotesize\ttfamily,
	breakatwhitespace=false,
	breaklines=true,
	captionpos=b,
	keepspaces=true,
	numbers=left,
	numbersep=5pt,
	showspaces=false,
	showstringspaces=false,
	showtabs=false,
	tabsize=2,
	frame=single,
	rulecolor=\color{black},
	framerule=0.3pt,
	keywordstyle=\color{blue},
	commentstyle=\color{green},
	stringstyle=\color{red},
	morekeywords={
		False, None, True, and, as, assert, break, class, continue,
		def, del, elif, else, except, finally, for, from, global,
		if, import, in, is, lambda, nonlocal, not, or, pass,
		raise, return, try, while, with, yield,
	},
	emph={update},
	emphstyle=\color{magenta},
	xleftmargin=10pt,  %
	xrightmargin=5pt,
}

\begin{document}

\title{Efficient Data-Parallel Continual Learning with Asynchronous Distributed Rehearsal Buffers}

\author{\IEEEauthorblockN{Thomas Bouvier$^1$, Bogdan Nicolae$^2$, Hugo Chaugier$^1$, Alexandru Costan$^1$, Ian Foster$^2$, Gabriel Antoniu$^1$}
\linebreak
\IEEEauthorblockA{\textit{$^1$University of Rennes, Inria, CNRS, IRISA, $^2$Argonne National Laboratory}}
\linebreak
\IEEEauthorblockA{
\{thomas.bouvier,hugo.chaugier,alexandru.costan,gabriel.antoniu\}@inria.fr,\{bnicolae,foster\}@anl.gov
}
}

\maketitle

\begin{abstract}
Deep learning has emerged as a powerful method for extracting valuable information
from large volumes of data. However, when new training data
arrives continuously (i.e., is not fully available
from the beginning), incremental training suffers from catastrophic forgetting
(i.e., new patterns are reinforced at the expense of previously acquired knowledge). 
Training from scratch each time new training data becomes
available would result in extremely long training times and massive data accumulation. 
Rehearsal-based continual learning has shown promise for addressing the catastrophic forgetting challenge, but research to date has not addressed performance and scalability. 
To fill this gap, we propose an approach based on a distributed rehearsal buffer that efficiently complements data-parallel training on multiple GPUs, allowing us to achieve short runtime and scalability while retaining high accuracy. It leverages a set of buffers (local to each GPU) and uses several asynchronous techniques for updating  these local buffers  in an embarrassingly parallel fashion, all while handling the communication overheads necessary to augment input mini-batches (groups of training samples fed to the model) using unbiased, global sampling. 
In this paper we explore the benefits of this approach for classification models. We run extensive experiments on up to 128 GPUs of the ThetaGPU supercomputer to compare our approach with baselines representative of training-from-scratch (the upper bound in terms of accuracy) and incremental training (the lower bound). Results show that rehearsal-based continual learning achieves a top-5 classification accuracy close to the upper bound, while simultaneously exhibiting a runtime close to the lower bound.


\end{abstract}

\section{Introduction}
\label{sec:introduction}
Deep learning (DL) models are rapidly gaining traction both in
industry and scientific computing 
in many areas, including speech and vision,
climate science, %
cancer research, %
to name a few~\cite{ALAM2020302,osti_1507546}. %

As data sizes and pattern complexity keep increasing,
DL models capable of learning such data patterns have evolved from
all perspectives: size (number of parameters), depth (number of
layers/tensors), and structure (directed graphs that feature divergent
branches, fork-join, etc.). Despite increasing convergence between DL and
HPC~\cite{HuertaKDBGKKKKM20}, which has led to the adoption of various
parallelization techniques~\cite{ben2019demystifying} (data-parallel,
model parallel, hybrid), the training of DL models remains a
time-consuming and resource-intensive task. 
Indeed, the amount of compute used in the largest AI training runs has doubled every 3.4 months since 2012.

DL models are typically trained on large, many-GPU ("HPC") systems that have access to
all training data from the beginning (e.g., from a parallel file system), by using an iterative optimization technique
(e.g., stochastic gradient descent) to revisit the training data repeatedly until convergence. 
However, DL applications increasingly need to be trained with
unbounded datasets that are updated frequently.
For example, scientific applications using experimental devices such
as sensors
need to quickly analyze the experimental data in
order to steer an ongoing experiment (e.g., 
trigger an automated decision).
In this case, repeatedly retraining the model
from scratch as new samples arrive is not an option: as training data keeps
accumulating, this would take increasingly longer and consume more resources (GPU hours, storage space),
leading to both prohibitive runtimes as well as inefficient resource usage.

One approach to this problem is to train the DL model incrementally (i.e., the training proceeds with
relatively inexpensive updates to the model's parameters based on just the new data samples).
If data increments are small, such an approach achieves high performance and low resource utilization. 
Unfortunately, it can also cause the accuracy of the DL model to deteriorate quickly---a phenomenon known
as \emph{catastrophic forgetting}~\cite{mccloskey1989catastrophic}. 
Specifically, the training introduces a bias in favor of new samples, effectively causing the model to reinforce recent patterns at the expense of previously acquired knowledge. Increased differences between the distributions of the old
vs.\ new training data amplifies the bias, often to the point where 
a single pass over the new training data is enough to erase most, if not all, of the
patterns learned previously.

Thus, we are faced with the challenge of avoiding catastrophic forgetting efficiently.
We aim to achieve an accuracy close
to the one achieved by retraining the DL model from scratch, but we also aim to achieve high performance, scalability,
and low resource utilization just like incremental training. 
To address this trade-off, \emph{continual learning} (CL) is gaining
popularity in the machine learning community~\cite{%
hadsell2020embracing}. In a broad sense, CL
mitigates catastrophic forgetting by complementing incremental
training with a strategy to reinforce patterns seen earlier.

Proposed CL strategies include
\emph{rehearsing} historic training samples, co-training a generative
DL model that can mimic old patterns by generating new samples on demand, and 
regularization (i.e., rules that constrain DL model
parameter updates to prevent catastrophic forgetting), among others.
We focus here on \textbf{continual learning based on
rehearsal}. With this strategy, historic training samples that are
representative of patterns seen earlier are retained in a limited-size rehearsal
buffer. 
Small subsets of incoming training samples 
(called \emph{mini-batches}) 
are then \emph{augmented} 
to include additional samples from the rehearsal
buffer. Finally, the rehearsal
buffer is updated by replacing some of its samples with newer ones.
A benefit of this CL strategy is that it requires no modifications to either the DL model architecture or the training process.
In contrast, other CL approaches require different hyperparameters, additional code to implement regularization, and/or additional 
generative DL models. 

Prior work on rehearsal-based CL~\cite{titsias2019functional,pan2020continual,mirzadeh2020understanding} has employed a single rehearsal buffer, with the goal of leveraging a single GPU.
Here, we tackle the problem of enabling high-performance, scalable,
and resource-efficient rehearsal-based CL on \textbf{multiple GPUs}. 
\emph{Data-parallel} training is commonly used to reduce training time. 
In this approach, the DL model is replicated on multiple GPUs (on the same
or different compute nodes) and each DL model replica is trained with a different data shard,
with the gradients computed by different replicas averaged periodically to keep the replicas in sync. 

Efficient continual learning based on rehearsal that delivers high
performance, scalability, and low resource utilization in combination
with data-parallel training is difficult for two reasons: (1) the cost of managing a
rehearsal buffer under concurrency (mini-batch augmentations and
constant updates) is significant, and (2) efficient data-parallel training
requires the instantiation of multiple independent rehearsal buffers, one
per DL model replica, thus limiting the possible combinations for
mini-batch augmentations (i.e., reducing both their diversity and quality).
To address these challenges, we propose the use of a \textbf{distributed
rehearsal buffer}: it focuses on how to minimize the
overheads involved by the rehearsal buffer management while retaining the
quality of mini-batch augmentations under data-parallel training.
We summarize our contributions as follows:

\begin{itemize}
\item We define the concept of rehearsal buffers to
address continual learning, and introduce \textbf{extensions to leverage them for data-parallel training} (Section~\ref{sec:buffer}).
\item We introduce key design principles such as \textbf{asynchronous
techniques to hide the overhead of managing rehearsal buffers} and to enable
a full spectrum of combinations for \textbf{mini-batch augmentations}. We achieve
this by sampling the rehearsal buffers of remote DL model replicas using
low-overhead, RDMA-aware, all-to-all communication patterns
(Section~\ref{sec:async}).
\item We leverage these design principles to implement
a \textbf{distributed rehearsal buffer prototype that we integrated
with PyTorch}, a popular AI runtime (Section~\ref{sec:implementation}).
\item We report on \textbf{extensive experiments on 128 GPUs of the ANL's ThetaGPU supercomputer}
with three different models (ResNet-50, ResNet-18, GhostNet-50) and four tasks derived from the ImageNet-1K dataset. \textbf{Note: We specifically focus on training classification models (models for generative AI are out of the scope of this paper).}
\item In the best case with ResNet-50, we show that our method can improve the top-5 evaluation accuracy from 23.1\% to 80.55\% compared with incremental training, with just a small runtime increase (Section~\ref{sec:evaluation}).
\end{itemize}

\section{Background and Problem Statement}
\label{sec:background}
In this section we revisit several key DL concepts to set the context for our work.

{\bf Basics of Deep Learning Training:} DL is an iterative process: starting with an initial set of weights $w$ chosen randomly, the training data is visited multiple times to update $w$.
Each full visit is called an \emph{epoch}, and during each epoch the training
data is shuffled and split into \emph{mini-batches}. Each mini-batch is processed
in an iteration that involves a \emph{forward pass} to predict the output, and a \emph{backward pass} that calculates the intermediate gradients corresponding to the differences between the output and the ground truth, which are then used to update $w$.

{\bf Data Parallelism:} a typical optimization used in practice is to create multiple DL model replicas on different GPUs, each of which is trained at the same time on a different shard of the training data~\cite{ben2019demystifying}, effectively reducing the number of iterations
in an epoch (hence called \emph{data-parallelism}). In this case, the forward
and backward pass can run independently, except that after each backward pass, 
the gradients computed by all replicas are averaged (by using a collective communication pattern such as \emph{all-reduce}) before adjusting $w$.  It ensures that the DL model replicas always apply the same updates on $w$ and are thus in sync.

{\bf Catastrophic Forgetting:}
Although efficient on static training data, the iterative
process does not perform well when new training data arrives over time. In this case, if we continue training the DL model using
only mini-batches from the new training data (called \emph{incremental training}), the DL model will drift in the direction of the new training data. This phenomenon is known as \textit{catastrophic forgetting}. It echoes the more general plasticity-stability dilemma~\cite{mermillod2013stability}, where (1) plasticity refers to the ability of the model to learn concepts in the current task, and (2) stability refers to its ability to preserve knowledge acquired in previous tasks.

{\bf Task-incremental CL vs.\ Class-incremental CL:}
continual learning (CL) aims to enable refinement of DL models
using continuously arriving new training data while 
mitigating catastrophic forgetting, such as to preserve the knowledge gained during the previous training. In the case 
of classification problems, we can either assume that
the output classes remain fixed (i.e., new training samples
belong to one of the pre-determined classes) or that they
can change (i.e., new training samples may introduce
new classes). The former is called \emph{task-incremental} 
while the latter is called \emph{class-incremental} continual learning~\cite{farquhar2018towards}. In this paper, we focus 
on the latter, which is the most general and difficult case. 
If the problem solved is not a classification but involves a generative DL model, this can be reduced to the case of task-incremental continual learning (i.e., the meaning of the output 
is fixed and does not change).

{\bf Streaming CL vs.\ Batched CL:}
another important aspect is how we reason about the increments:
we can either assume the new training data continuously
arrives from a stream and can be visited only once before
it is discarded (called \emph{streaming} CL or single
epoch CL~\cite{hu2021one}), or the new training data 
arrives in batches that can stored and revisited over multiple epochs (called \emph{batched} CL). The latter is more
common in practice, hence we focus on it in this work.

{\bf Problem Statement:}
As mentioned in Section~\ref{sec:introduction}, retraining the model from scratch on all previously accumulated
data is not feasible, because each epoch would contain more and more mini-batches as more tasks are being learnt.
This would cause long delays until the DL model is ready for inferences after each task. Furthermore, this would
cause an explosion of storage space needed to retain the history of all training
samples. Our goal is to devise scalable CL techniques that retain a classification accuracy close to the
train-from-scratch approach (which is the upper bound), while simultaneously achieving a runtime
close to incremental training (which is the lower bound). The main
research question we aim to answer is \textbf{how to combine rehearsal-based 
continual learning with
data-parallel training} in order to achieve this goal. 

\section{Related Work}
\label{sec:related-work}

\textbf{Experience Replay} (that we refer to as \textit{rehearsal}) is a simple continual learning technique in which the model knowledge is reinforced by replaying samples from previous tasks~\cite{ratcliff1990connectionist}%
. These methods selectively store previously encountered raw data samples, called \textit{representatives} (sometimes referred to as \textit{exemplars}), into a \textit{rehearsal buffer}, which
is used to \emph{augment} the mini-batches of new training tasks.
The augmentation involves appending a fixed number of representatives to each mini-batch corresponding to the new 
training data in order to obtain a large mini-batch that mixes
new and old training samples. The advantage of this approach
is that it can mitigate catastrophic forgetting transparently
~\cite{balaji2020effectiveness}, without the need to change existing training methods. This claim is supported by works that not only emphasize its efficacy compared to alternative methods~\cite{rolnick2019experience}, but also propose diverse extensions to enhance its performance~\cite{buzzega2021rethinking}. %
HAL~\cite{chaudhry2020using} complements Experience Replay with regularization to align the model responses with data points encoding classes encountered in previous tasks.
DER (Dark Experience Replay) and DER++~\cite{buzzega2020dark} demonstrate that replaying model responses instead of data labels (or doing both) yields to a better achieved accuracy than Experience Replay alone. eXtended-DER~\cite{boschini2022class} takes an extra step over previous methods by preparing future classification heads to accommodate future classes.

\textbf{Data Management Techniques for Training Data.} Reading the training data directly 
from a shared repository (such as a parallel file system) has been shown to introduce
significant bottlenecks~\cite{AI-FlexScience22}.
DeepIO~\cite{zhu2018entropy} uses a partitioned caching technique
for data-parallel training, relying heavily on RDMA for high performance I/O.
DIESEL~\cite{wang2020diesel} deploys a distributed cache across compute nodes to handle multiple
DL training instances sharing the same training data. 
MinIO~\cite{mohan2020analyzing} focuses
on eviction-free caching of training data, which has low overheads and is easy to implement but may
lead to higher miss rate. NoPFS~\cite{dryden2021clairvoyant} introduces a performance model that can
leverage multi-level node-local storage for distributed caching of training samples. 
Lobster~\cite{LobsterAI-ICPP22} further refines this approach by optimizing cache evictions 
and by enabling load balancing 
in the
data pipeline. Such approaches optimize the data pipeline
and complement our proposal.

\textbf{Positioning.} In this work, we propose
asynchronous data management techniques that enable
the design and implementation of a scalable distributed rehearsal buffer abstraction, which is
instrumental in enabling continual learning to
take advantage of data-parallel techniques. To our best
knowledge, we are the first to explore this direction.

\section{Contribution: Distributed Rehearsal Buffers}
\label{sec:buffer}

\newlength\mylengtha
\newlength\mylengthb
\settowidth\mylengtha{$A \subsetneq B$}
\setlength\mylengthb{\dimexpr\columnwidth-\mylengtha-6\tabcolsep\relax}
\begin{table}[t]
    \caption{Continual Learning notation}
    \centering %
    \begin{xtabular}{p{\mylengtha} p{\mylengthb}}
        \toprule
        $T$ & number of CL tasks\\
        $K$ & number of classes\\
        $\mathcal{B}$ & distributed rehearsal buffer\\
        $\mathcal{B}_n$ & local rehearsal buffer for process $n$\\
        $R^i_n$ & subset of $\mathcal{B}_n$ containing representatives of class $i$\\
        $c$ & number of candidates per mini-batch\\
        $b$ & mini-batch size (number of samples per mini-batch)\\
        $r$ & number of representatives added to augmented mini-batches\\
        \bottomrule
    \end{xtabular}
\end{table}

In this section we discuss the key design principles that are at the foundation of our proposal.

\subsection{Distributed Rehearsal Buffer}
\label{sec:rehearsal-buffer}

In a basic version of rehearsal, a buffer $\mathcal{B}$ stores \emph{representative} data samples
from previous tasks. Every class $i$ observed so far is attached to its own rehearsal
buffer $R^i \in \mathcal{B}$. At each iteration, $r$ representatives from $\mathcal{B}$ 
are used to \emph{augment} the incoming mini-batch
$m$ of size $b$, such that we obtain a larger mini-batch of size $b+r$ mixing representatives
and new training samples. This new mini-batch is an \emph{augmented mini-batch}. After training with the augmented mini-batch, 
$c$ training samples, called \textit{candidates}, are selected from mini-batch $m$
to be inserted into the relevant buffer $R^i$. If any of the $R^i$ buffers is full,
then the new candidates replace old representatives as needed (e.g., at random or using
a different strategy). This process ensures that each $R^i$ buffer remains up-to-date
at fine granularity (i.e. after each iteration), holding representatives of both the current and all previous tasks. 

\begin{figure}[t]
  \centering
  \includegraphics[width=\columnwidth]{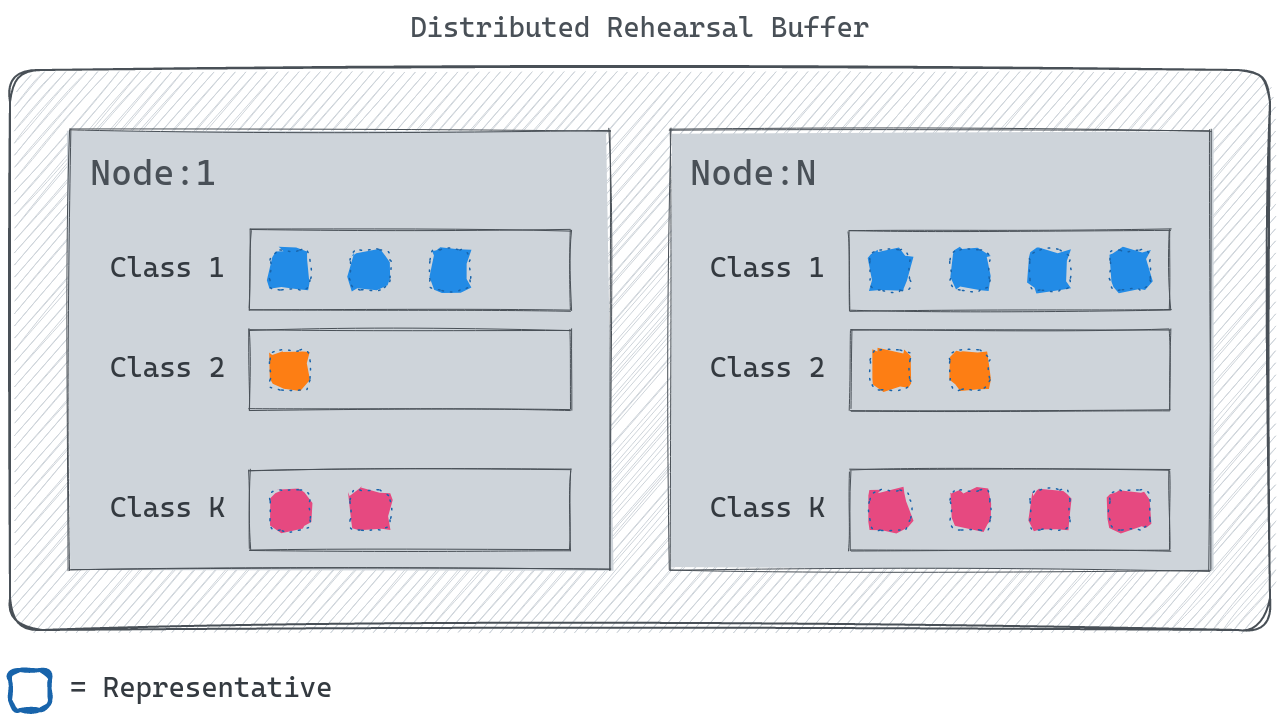}
  \caption{For every process $n$, a rehearsal buffer $\mathcal{B}_n$ contains representatives from the classes seen so far. The distributed rehearsal buffer $\mathcal{B}$ contains representatives from the $K$ classes.}
  \label{fig:distributed-buffer}
  \vspace{-10pt}
\end{figure}

Starting from this basic version, we propose to design a \emph{distributed rehearsal
buffer} that can be used with data-parallel training. In our case, the training
uses $N$ distributed processes (each attached to a dedicated GPU). Each process maintains its own rehearsal buffer $\mathcal{B}_n$. Thus, we can leverage the aggregated spare memory provided by a large number of compute nodes to store more representatives compared with a single centralized buffer.
Conceptually, the disjoint union of local rehearsal buffers $\mathcal{B}_n$ can be seen as a single distributed rehearsal buffer $\mathcal{B}$ as depicted in Figure~\ref{fig:distributed-buffer}:

$$\mathcal{B}=\bigsqcup\limits_{n=0}^{N}\bigsqcup\limits_{i=0}^{K}R^{i}_{n}=\bigsqcup\limits_{n=0}^{N}\mathcal{B}_n$$

Assume each process can spare up to $S_{max}$ local memory for storing 
$\mathcal{B}_n$. Given increasing DL model sizes, the spare host and GPU memory 
is under pressure, thus $S_{max}$ is limited. On the other hand, we need to
divide $S_{max}$ evenly between the classes to
avoid a bias in the selection of the representatives. Therefore, each $R^i_n$ can 
grow up to a size of $S_{max} / K$, which means with increasing number of classes $K$, 
each buffer $R^i_n$ shrinks. However, by using a distributed rehearsal buffer,
each $R^i_n$ scales with the number of processes to a size of
$|R^i_n|_{max} = N \times S_{max} / K$, which increases the number of representatives
per class and therefore the diversity and quality of the mini-batch augmentation.
This complements data-parallel training well, since data-parallel training
improves performance and scalability, not the quality of the results.

\subsection{Selection and Eviction Policies}
\label{sec:policies}

Since the rehearsal buffer $\mathcal{B}$ is smaller than the dataset $D$, we 
are concerned about selection and eviction policies for managing the distributed 
rehearsal buffer. One approach to populate the local rehearsal buffers is to
select candidate samples from incoming mini-batches at random. To this end, we propose
Algorithm~\ref{alg:naive-selection}, which is executed by each process $n$ at every 
training iteration. Specifically, we pass the current mini-batch $m_n$ of size $b$.
Every sample of $m_n$ has a $c / b$ probability to be pushed into the buffer $R^i_n$
corresponding to the class $i$. As such, $c$ acts like an update rate: i.e. the higher
the $c$, the more often representatives are renewed in rehearsal buffer $\mathcal{B}_n$. 
This approach has been implemented in the \textit{Naive Incremental Learning} (NIL)
algorithm~\cite{munoz2020incremental} and demonstrates low computational complexity.

\begin{figure}[t]
  \centering
  \includegraphics[width=\columnwidth]{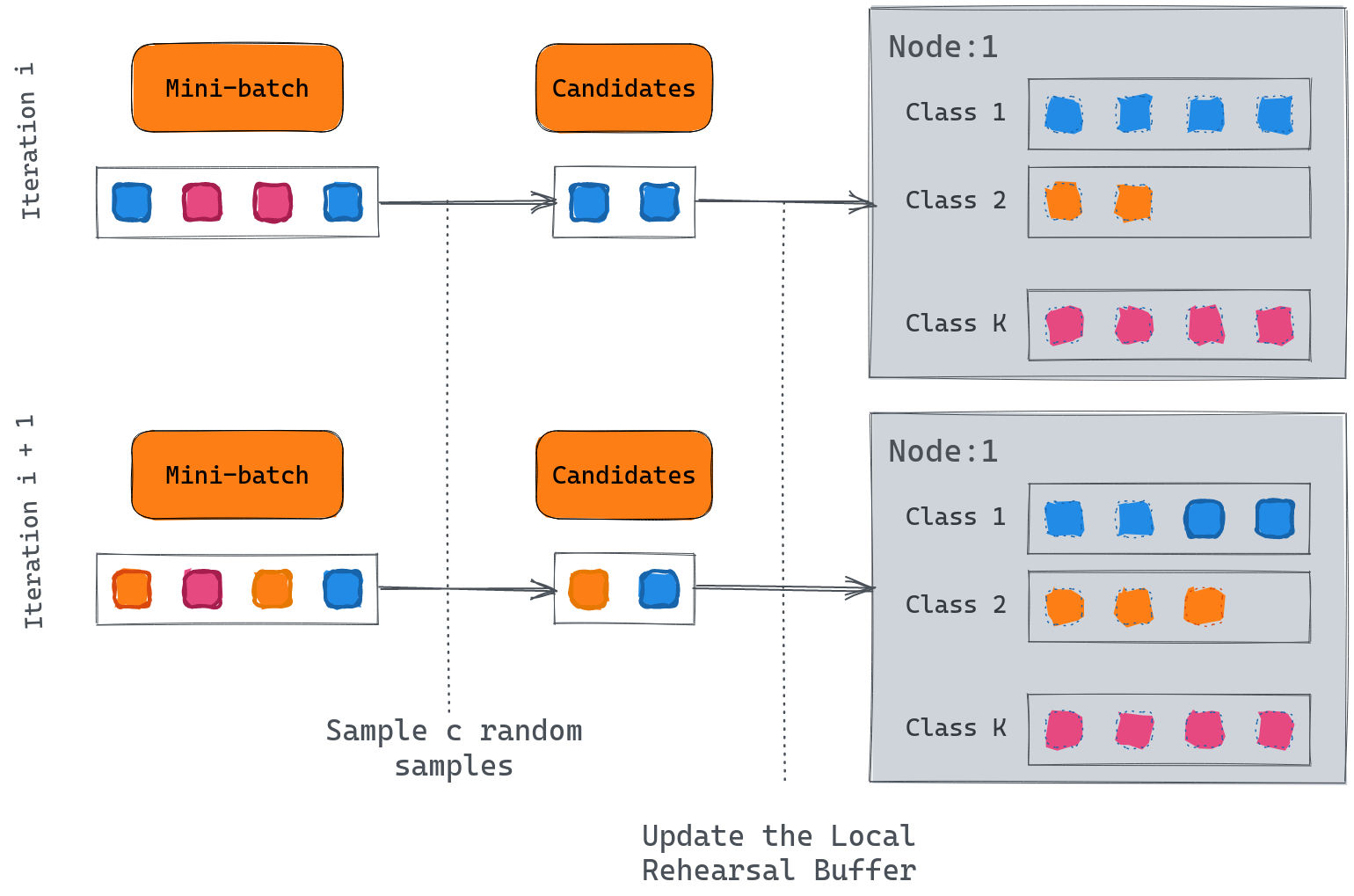}
  \caption{For a given process $n$, $c$ candidates from the incoming mini-batch are sampled and used to populate $\mathcal{B}_n$. If the buffer for class $i$ is full, representatives from $R^i_n$ are replaced at random. The figure depicts the rehearsal buffer $\mathcal{B}_n$ state for two subsequent iterations for $c = 2$.}
  \label{fig:candidates}
\end{figure}

\begin{algorithm}
    \small
    \caption{Rehearsal buffer updates with new candidates for each process $n$}
    \label{alg:naive-selection}
    \DontPrintSemicolon
    \SetKwFunction{Update}{update\_buffer}
    \SetKwProg{Fn}{Function}{:}{}
    \Fn{\Update{$m$}}{
        $C \gets$ select $c$ random candidates from mini-batch $m$ \;
        \For{$c \in C$} {
            \If{$|R^i_n|$ >= $|R^i_n|_{max}$} {
               replace a random representative from $R^i_n$ with $c$ \;
            } \Else {
                append $c$ to $R^i_n$ \;  
            }
        }        
    }
\end{algorithm}

Since representatives are distributed among $R^i_n$ according to their class labels, the competition between new candidates and stored representatives is done by class. Thus, candidate samples belonging to a specific class compete against the existing representatives of the same class. As depicted in Figure~\ref{fig:candidates}, a candidate sample of class $i$ replaces a random representative in $R^i_n$ if the latter is full.  Our random selection policy means that each training sample of a given class has the same probability of being replaced, regardless of whether it is a recent or old sample. Thus, we \emph{independently} obtain a good mix of old and new training samples in 
each buffer $R^i_n$. This approach both increases the quality of the augmentations and forms an embarrassingly parallel pattern that is easy to implement
and that has a low performance overhead. The selection and eviction policies introduced here are operating at the process level. When working with nonuniform sample distributions across data shards, load balancing strategies could fill remote buffers.

\subsection{Global Mini-Batch Augmentation using RDMA-enabled Distributed Sampling of Representatives}
\label{sec:global-sampling}

\begin{figure}[t]
  \centering
  \includegraphics[width=\columnwidth]{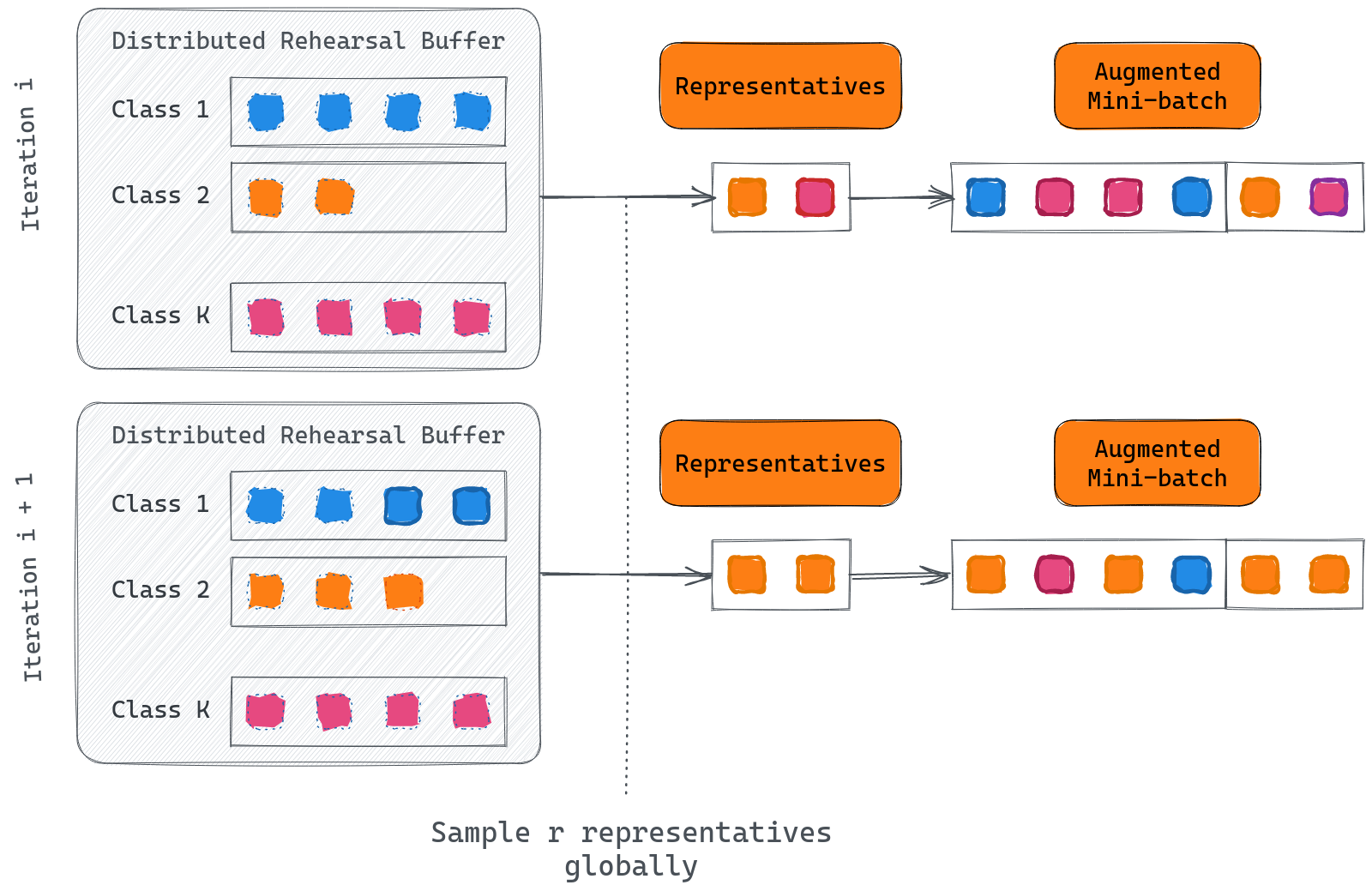}
  \caption{On a given process $n$, every incoming mini-batch is augmented with $r$ representatives sampled randomly and without replacement from the distributed rehearsal buffer $\mathcal{B}$. Here, $r = 2$ on two subsequent iterations. Sampling from $\mathcal{B}$ introduces communication between the $N$ distributed processes. }
  \label{fig:augmentations}
\end{figure}

Experience Replay consists in interleaving representatives with the current mini-batch $m$ to build a new augmented mini-batch $m'$. As depicted in Figure~\ref{fig:augmentations}, at every training iteration, $r$ representatives are sampled without replacement from $\mathcal{B}$ to assemble $m'$, whose size is $b+r$. We call this operation \emph{mini-batch augmentation}. Existing research has shown that uniform sampling from a rehearsal buffer is effective in many cases~\cite{munoz2020incremental,
balaji2020effectiveness}, while demonstrating no additional computational complexity. Thus, we adopt the same principle in our proposal.

With a distributed rehearsal buffer $\mathcal{B}$, each process $n$ needs to sample
$r$ representatives concurrently with the other processes. To this end, we could
simply adopt a naive embarrassingly-parallel strategy that chooses the $r$ 
representatives of each process $n$ from the local rehearsal buffer $\mathcal{B}_n$. 
Although highly efficient and easy to implement, such a strategy limits the 
number of combinations possible for the selection of the $r$ representatives
relative to the global rehearsal buffer $\mathcal{B}$, which reduces the diversity 
and the quality of the augmentations. This effect is similar to the bias introduced
by sharding for data-parallel training (as discussed in Section~\ref{sec:background}).
As a consequence, we need to provide a fair sampling that gives every training
sample in $\mathcal{B}$, regardless of its location, an equal opportunity to be
selected among the $r$ representatives of each process. This is a difficult
challenge for several reasons: (1) competition for network bandwidth, since
many processes sharing the same compute node need to transfer training samples 
from remote rehearsal buffers at the same time; (2) difficult all-to-all 
communication patterns, since each process needs to access the rehearsal buffers
of every other process; (3) low latency requirements, since each process needs
to access a small number of training samples from each remote rehearsal buffer.

To address this challenge, we leverage two technologies commonly used
in HPC. First, we propose to pin the space reserved for each local rehearsal 
buffer $B_n$ into the memory of the compute node hosting process $n$. Then,
we expose the pinned memory for RDMA access. Using this approach, we enable
low-overhead, fine-grain access to the rehearsal buffer of each process
from every other process. Second, since the requests of the processes to
sample remote rehearsal buffers are not synchronized, we cannot simply
rely on existing patterns such as MPI all-to-all collective communication,
as this would introduce unnecessary delays. Therefore, we propose an RPC-based
communication pattern atop Mochi~\cite{Mochi20}, an HPC-oriented set of services
that provides low-overhead RDMA-enabled point-to-point RPCs. Specifically, we 
introduce several key concepts such as: (1) progressive assembly of 
augmented mini-batches using concurrent asynchronous RPCs, which hide the
remote access latency; (2) RPC consolidation to transfer the training samples 
in bulk from the same remote rehearsal buffer, reducing the number of
RPCs; (3) concurrency control based on fine-grain locking to guarantee
consistency and mitigate contention between updates to the rehearsal buffers 
and local/remote reads issued by augmentations. 

\subsection{Asynchronous Management of Rehearsal Buffers}
\label{sec:async}

Even with our proposed optimizations, the overheads of managing a distributed
rehearsal buffer may still be significant. Therefore, we also devise an asynchronous 
technique to hide these overheads, such that a training iteration can proceed without blocking every time that it needs to interact with
the distributed rehearsal buffer. 

To this end, we revisit the major steps  
of CL based on rehearsal and data-parallel training: \circled{1} prepare 
the augmented mini-batches, which involves global sampling from the distributed 
rehearsal buffer; \circled{2} update the distributed rehearsal buffer using
the new samples of the original mini-batches; \circled{3} perform a forward pass
with the augmented mini-batch as input data; \circled{4} perform a backward pass that averages
the gradients and updates the parameters $w$ of each DL model replica. One key
observation is that we start with an empty rehearsal buffer. Hence, for
the first training iteration we do not need to perform an augmentation. However,
after step \circled{2}, we can prepare an augmented mini-batch in advance for
the next training step. This applies for all subsequent iterations. 

\begin{figure}[t]
  \centering
  \includegraphics[width=\columnwidth]{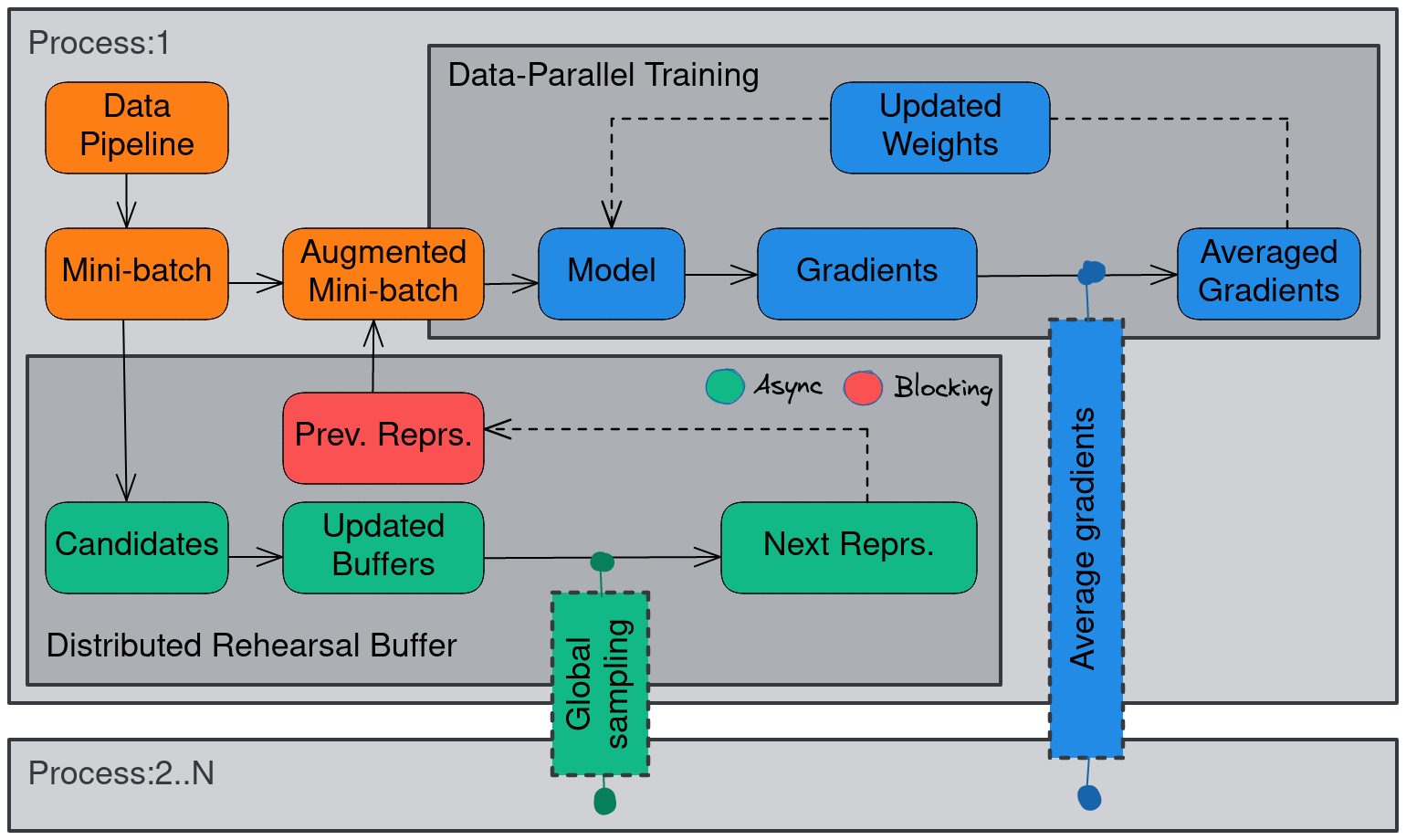}

  \caption{Asynchronous updates of the rehearsal buffers and global augmentations: 
  $r$ representatives sampled globally beginning with the previous iteration are
  used by the training loop to assemble an augmented mini-batch on each process
  $n$. Meanwhile, the distributed rehearsal buffer extracts candidates from the
  current mini-batch to update each $B_n$ locally, then collects
  the next $r$ representatives using global sampling.}
    \label{fig:flow}
\end{figure}

Therefore, we can use the following strategy: \circled{1} wait until $r$ 
representatives were collected asynchronously by global sampling started
during the previous iteration and concatenate them with the current 
mini-batch to obtain an augmented mini-batch; \circled{2}
start an asynchronous update of the distributed rehearsal buffer using
the original mini-batch, followed by asynchronous global sampling of the next $r$ 
representatives; perform the same steps \circled{3} and \circled{4} as above.
This process is illustrated in Figure~\ref{fig:flow}.

Using this approach, the communication and synchronization overheads related
to the management of the rehearsal buffer can be overlapped with the training steps.
Indeed, the training iterations only need to wait if the updates of the rehearsal
buffer and the global sampling cannot keep up with it and introduce a delay. 

It is important to note though that even in the case when the rehearsal buffer overhead 
can be fully absorbed asynchronously (i.e. no wait at step \circled{1}), the training
iteration operates with an augmented mini-batch of size $b+r$ (instead of the 
original size $b$). Thus, each training iteration is slowed down by a factor of $r/b$. 
This overhead is inherent to rehearsal-based CL and cannot be avoided. However, by 
fixing $r$ and hiding the rehearsal buffer management overheads through asynchronous
techniques, our approach can deliver performance levels close to the theoretical
lower bound at scale.

\section{Implementation Details}
\label{sec:implementation}
We implemented our approach as a high-performance, open-source C++ library~\cite{bouvier2021distributed} that offers convenient bindings for Python using~\emph{pybind11}.

There are multiple reasons for this choice:
(1) our approach requires low-overhead access to system-level resources, notably
RDMA-enabled RPCs, which is not available for Python; (2) even if bindings existed,
the overheads of interpreted languages are unacceptable in our case given the
need to provide consistency and manage multiple connections under concurrency; (3)
Python has limited support for multi-threaded concurrency due to the global
interpreter lock that allows only a single thread to run interpreted code. Nevertheless, the complexity of our proposal is completely hidden from end-users:
the distributed rehearsal buffer integrates seamlessly with the training loop
using a convenient \emph{update} primitive encapsulating all our contributions, illustrated in Listing~\ref{lst:training-loop}. 

\vspace{0.35cm}
\begin{lstlisting}[style=mystyle, caption={Example of a training loop integrating our proposal. The \emph{update} primitive (highlighted) waits until $r$ representatives were collected by the asynchronous global sampling, then updates the rehearsal buffer, and starts the next global sampling.}, label={lst:training-loop}]
for i in range(no_minibatches):
		m = DataPipeline.get_next_minibatch()
		r = RehearsalBuffer.update(m)
		m_a = concat(m, r)
		Model.train(m_a)
\end{lstlisting}

For the purpose of this work, we integrate our proposal with PyTorch %
and rely on Horovod~\cite{sergeev2018horovod} to enable data parallelism. We rely on NVIDIA DALI as the data pipeline that provides the original
mini-batches while overlapping with the training iterations.
Thanks to the encapsulation into a separate primitive, our
approach can be easily extended to support other AI runtimes (such as TensorFlow%
), data-parallel implementations or data pipelines.

To take advantage of high-performance, fine-grain parallelism, all operations are executed in a separate system pthreads and CUDA operations involving copies between GPUs and host memory are executed in a dedicated CUDA stream isolated from the Python frontend. We used Argobots (part of the Mochi framework~\cite{Mochi20}) for implementing a low-overhead, userspace thread pool that serves concurrent requests to update and read the training samples from rehearsal buffers. 
We used Thallium (also part of Mochi) to implement global sampling leveraging non-blocking RDMA-enabled RPCs.

\section{Evaluation}
\label{sec:evaluation}
In this paper we specifically focus on training classification models. We perform experiments on ANL's ThetaGPU supercomputer to study the benefits of our proposal with respect to both classification accuracy and training duration. Our evaluation seeks to answer the following questions:

\begin{itemize}
    \item How do parameters $r$ (representative count) and $|\mathcal{B}_n|$ (rehearsal size) impact achieved classification accuracy? %
    \item How much does accuracy degrade with CL, compared to the case where the model re-learns from scratch each time new data arrives?
    \item Do mini-batch augmentations increase training time?
    \item How much does training time increase, compared to incremental training?
    \item What is the memory cost of rehearsal-based learning?
\end{itemize}

\subsection{Experimental Setup and Methodology}

\textbf{Training Dataset:} we use the ImageNet-1K %
dataset, which is widely used in the image classification
community. We specifically use the variant with face-blurred images~\cite{yang2022study}, containing 1.2M training images split among 1000 object classes. Each class contains about 1300 training and 50 validation samples. We use standard data augmentations of random horizontal flips and crops resized at 224x224 pixels.

\textbf{Continual Learning Scenario:} we recall
that we focus on the class-incremental (``Class-IL'') scenario, in which there are clear and well-defined boundaries between the tasks to be learned (i.e. there is no overlap between classes of different tasks). We design a sequence of 4 disjoint tasks, each containing 250 classes from ImageNet. Each of them gets revisited 30 times (i.e. the model is trained for 30 epochs on every task), which corresponds to a total of 120 training epochs. The model can not revisit previous tasks.

\textbf{Learning Models:} to show that our rehearsal-based approach to CL is transparent with respect to the model, we use the 3 following convolutional networks and their corresponding configurations:

\begin{itemize}
    \item \textbf{ResNet-50}~\cite{he2016deep} is the standard with ImageNet. We use the SGD optimizer with a learning rate of 0.0125; a per-task learning rate increase on 5 warmup epochs as in~\cite{goyal2017accurate}; a gradual decay from 0.5 to 0.05 to 0.01 at epochs 21, 26, and 28, respectively; and a weight decay of 1e-5.
    \item \textbf{ResNet-18}~\cite{he2016deep} has roughly half the number of parameters of ResNet-50 and is thus faster to train (i.e. its minibatch processing time is shorter). We use the same hyperparameters as ResNet-50.
    \item \textbf{GhostNet-50}~\cite{han2020ghostnet} implements a different architecture 
    to minimize inference time on resource-constrained devices. We use the SGD optimizer with a learning rate of 0.01; the same warmup as ResNet's; the same schedule at epochs 15, 21, 28; and a weight decay of 1.5e-5.
\end{itemize}

We enable Automated Mixed Precision (AMP) introduced in~\cite{micikevicius2018mixed} to speed up the training.

\textbf{Scale:} we apply the linear scaling rule~\cite{goyal2017accurate} by multiplying the learning rate with the number of processes $N$. However, with an augmented mini-batch size set to $b+r=63$ training samples, which corresponds to a global batch size of $N \times 63$ in our data-parallel setting, the latter becomes greater that 8K with $N=128$. This requires further consideration to mitigate the instability introduced by such large batches~\cite{goyal2017accurate}. %
We do so by setting a maximum rate independent of the mini-batch size equal to $64$, as suggested theoretically in~\cite{bottou2018optimization}.

\textbf{Performance Metrics:} we report the top-5 accuracy achieved on the validation set to measure the model performance. Top-5 accuracy means any of the model's top 5 highest probability predictions is considered as correct. Let $a_{i, j}$ denote the top-5 evaluation accuracy on task $j$ using the model snapshot obtained at the end of task $i$. The accuracy (fraction of correct classifications) assessing the DL model performance on all previous tasks is defined as follows:

\begin{equation} \label{eq:1}
accuracy_T = \frac{1}{T}\sum_{j=1}^{T} a_{T, j}
\end{equation}

\begin{figure*}[ht]
  \centering
    \begin{subfigure}[b]{0.33\textwidth}
      \centering
      \includegraphics[width=\textwidth]{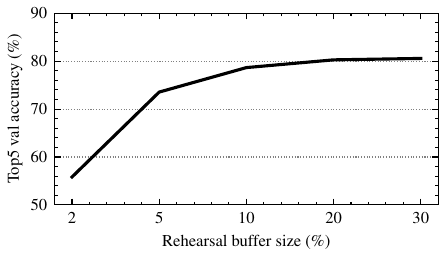}
      \caption{Accuracy w.r.t. different rehearsal buffer sizes $|\mathcal{B}|$ (percentage of the input dataset). Each data point is the average of the top-5 accuracy obtained on all previous tasks.}
      \label{fig:task_metrics_B}
    \end{subfigure}
     \hfill
     \begin{subfigure}[b]{0.66\textwidth}
         \centering
         \includegraphics[width=\textwidth]{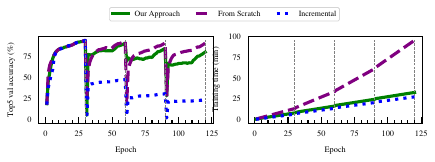}
         \caption{$|\mathcal{B}|=30\%$
         and $r=7$. Left: accuracy w.r.t. epoch number. Our rehearsal-based
         approach achieves a final accuracy of 80.55\%.
         Right: training time w.r.t. epoch number. Our approach induces a
         small runtime increase compared with incremental training, which
         stays linear.}
         \label{fig:resnet50}
     \end{subfigure}
     \caption{Top-5 accuracy for ResNet-50, 16 GPUs, ImageNet (4 tasks).}

\vskip -5pt
\end{figure*}

\textbf{Computing Environment:} we run our experiments on up to 16 nodes 
of ANL's ThetaGPU supercomputer (128 GPUs). Each node comprises eight
NVIDIA A100 GPUs (40 GiB HBM), two AMD Rome CPUs and NVIDIA Mellanox ConnectX-6 interconnect technology. We use the following
software environment: Python 3.10, PyTorch 1.13.1, Horovod 0.28.1, CUDA 11.4,
NVIDIA DALI 1.27.0, OpenMPI 4.1.4, Mercury 3.3 as well as libfabric 1.16
compiled with CUDA support.

\subsection{Impact of the Rehearsal Buffer Size on Accuracy}
\label{sec:impact-buffer-size}

As detailed in Section \ref{sec:rehearsal-buffer}, distributing the training
across $N$ processes allows to leverage the aggregated memory to store more
representatives in the rehearsal buffer $|\mathcal{B}| = N \times |\mathcal{B}_n|$.
Sampling representatives globally allows to distribute a
certain percentage of the input dataset over all processes (e.g., storing
10\% of the input dataset means in practice storing $10\%/N$ of the data per process).
To showcase the effect of different rehearsal buffer sizes on the accuracy, we
vary $|\mathcal{B}|$ from 2.5\%, 5\%, 10\%, 20\%, to 30\% of the total number
of ImageNet data samples (1.2M images). These values correspond respectively to
1.93 GB, 3.85 GB, 7.71 GB, 15.41 GB and 23.12 GB of raw data stored in the
aggregated memory.

We measure the performance of our approach with different rehearsal buffer sizes
by applying Equation~\ref{eq:1} once at epoch 120 (end of the training), in order
to evaluate the DL model on all previous tasks i.e. on all the classes seen until
then. We consider only ResNet-50 for this study, and run these experiments on 2 nodes (16 GPUs). We report the results in
Figure~\ref{fig:task_metrics_B}. As expected, the larger the rehearsal buffer size
$|\mathcal{B}|$, the better the diversity among stored representatives. As a result,
the model forgets less knowledge acquired in previous tasks, resulting in a higher
final accuracy. In our setting, storing 30\% of the input data samples as
representatives yields to a final top-5 accuracy of 80.55\%, which is significantly
better than the accuracy achieved with $|\mathcal{B}| = 2.5\%$ (55.83\%). We emphasize
that storing 30\% of ImageNet samples amounts to storing 1.45 GB of raw data per
process (with $N = 16$), which is only a fraction of the memory available on typical
HPC systems (512~GB of host memory per compute node).

\subsection{Impact of Other Rehearsal-related Hyperparameters}
\label{sec:impact-representative-count}

Parameter $c$ (introduced in Section~\ref{sec:policies}) is less relevant in
class-incremental scenarios, as: 1) classes from different tasks are disjoint,
and 2) the competition to populate the buffer is done per class. As a result,
representatives from previous tasks never get evicted under this setting. We
set $c=14$, which in our experimental setup only impacts the renewal rate of representatives from the current task.

Parameter $r$ (introduced in Section~\ref{sec:global-sampling}) has a direct
impact on the balance between plasticity and stability, where the model should
be both plastic enough to learn new concepts, and stable enough to retain knowledge. Mixing too many representatives with incoming mini-batches decreases the ability of the DL model to learn the current task, resulting in a degraded accuracy. A larger value for $r$ favors stability over plasticity. Authors in~\cite{munoz2020incremental} set $r$ to 15\% of the mini-batch size $b$: we adopt a similar ratio, setting $b=56$ and $r=7$.

\subsection{Comparison with Baseline Approaches}

We apply the insights obtained in the experiments detailed in Sections~\ref{sec:impact-buffer-size} and \ref{sec:impact-representative-count}, and we set $|\mathcal{B}|=30\%$ and $r=7$ to achieve high accuracy in the remainder of this paper. The following baselines instantiate models without any regularization or rehearsal:

\begin{itemize}
    \item \textbf{Incremental training}: updates the model with the training data
    corresponding to a single task, one at a time. No training samples of any previous
    tasks are revisited.
    \item \textbf{Training from scratch}: re-trains the model from scratch at every
    new task, using all accumulated training samples of both the new task and the
    previous tasks.
\end{itemize}

We consider ResNet-50 for this study. In Figure~\ref{fig:resnet50}, the top-5
evaluation accuracy achieved by rehearsal (80.55\%) greatly outperforms the incremental
training baseline. The latter suffers from catastrophic forgetting and is regarded as
the lower bound accuracy-wise (23.3\%). On the opposite, training from scratch as new
data arrives is regarded as the upper bound (91\%), only about 10.5\% above the
accuracy achieved with our rehearsal-based approach.

\begin{figure*}[t]
  \centering
  \includegraphics[width=0.80\textwidth]{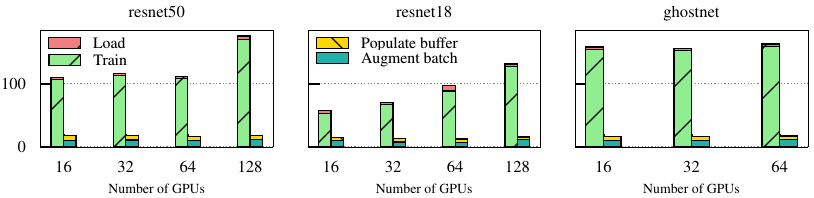}  
  \caption{Time breakdown (ms) for the training loop
  and rehearsal buffer management, for each of the three models and for different numbers of GPUs, each averaged across 35 minibatches.}
  \label{fig:breakdown}
\end{figure*}

In Figure~\ref{fig:resnet50}, we observe that incremental training has the shortest runtime as no task
gets revisited (lower bound). On the other hand, the duration of training from scratch increases
quadratically with the number of tasks to learn $T$. This is noticeable as a large gap
between the two approaches as the number of tasks increases. Just like incremental
training, our rehearsal-based approach exhibits a linear runtime with just a slight increase
proportional to the $r$ additional samples added to the minibatch. We demonstrate in the next section
that this overhead is not introduced by the rehearsal buffer management itself. Thus, we
conclude that our approach combines the best of both baselines: in terms of accuracy,
it is close to the train-from-scratch approach, while simultaneously nearing incremental training in terms
of performance.

\subsection{Rehearsal Buffer Management Breakdown}
\label{sec:augmentation-breakdown}

In Figure~\ref{fig:breakdown} we examine the time taken for the individual operations within a training iteration.
This study allows us 
to understand how well our approach overlaps the rehearsal buffer
management with the actual training process.

Specifically, we measure the time taken to obtain a new minibatch from
DALI (denoted \textit{Load}), which itself uses an asynchronous data
pipeline that prefetches and shuffles the training data. Then, we measure
the duration of the forward and backward passes as reported by PyTorch
(denoted \textit{Train}). The time taken for \textit{Load} followed by \textit{Train}
is the lowest possible overhead perceived by the application; this time is
represented by the stacked bars on the left of each of the 11 pairs of data bars in Figure~\ref{fig:breakdown}. In the background, our approach
handles updates to the individual rehearsal buffers (denoted \textit{Populate
buffer}), the distributed sampling of the remote rehearsal buffers, and 
the mini-batch augmentation (denoted \textit{Augment batch}); this time
is represented by the right-hand stacked bars in the figure. As long as the stacked bars
on the right are lower than those on the left, our approach will not cause
the training iterations to wait for the augmented mini-batches. This
means there is a full overlap and the rehearsal buffer management
is completely hidden in the background thanks to our asynchronous techniques.

Indeed, we observe that this condition holds for all models and all scales used in our experiments. Furthermore, the total overhead
of our approach is just a fraction of the \textit{Load} and \textit{Train}
overheads. Since the \textit{Train} overhead dominates (thanks to
DALI's asynchronous data pipeline), we conclude that there is a large 
margin left to optimize the forward and backward passes
without reducing the effectiveness of our approach. 

Moreover, another interesting effect is visible: we cannot simply reduce
the duration of the forward pass and backward pass at scale by optimizing 
the computations: when we switch from ResNet-50 to ResNet-18, which is 
significantly less computationally expensive to train, the duration
of \textit{Train} increases because all-reduce gradient reductions 
are expensive and begin to stall the computations. Thus, based on the observed
trends, we hypothesize that our approach remains effective at scale even in 
extreme cases of computationally trivial models.

\subsection{Scalability}

\begin{figure*}[ht]
  \centering
    \begin{subfigure}[b]{\textwidth}
      \centering
      \includegraphics[width=\textwidth]{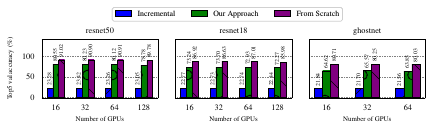}
      \vspace*{-7mm}
      \caption{Final top5 evaluation accuracy w.r.t.
      the number of processes (GPUs) $N$.}
      \label{fig:scaling-accuracy}
    \end{subfigure}
    \hfill
    \begin{subfigure}[b]{\textwidth}
      \centering
      \includegraphics[width=\textwidth]{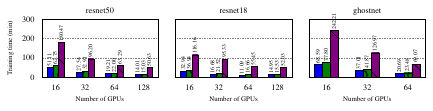}
      \vspace*{-7mm}
      \caption{Runtime w.r.t. the number of processes (GPUs) $N$. }
      \label{fig:scaling-time}
    \end{subfigure}
    \caption{Accuracy and runtime,  $|\mathcal{B}|=30\%$, $b=56$ and $r=7$ for all 3 models. For ResNet-50, colors match those in Fig.~\ref{fig:resnet50}.}
\end{figure*}

We study the scalability of our approach for all three models compared with 
the two baselines for an increasing number of data-parallel processes (GPUs).
Specifically, we measure the final evaluation top-5 accuracy in Figure~\ref{fig:scaling-accuracy}, where Equation~\ref{eq:1} is
applied once at epoch 120. We also measure the overall runtime to train all tasks and depict it in Figure~\ref{fig:scaling-time}.

All three approaches retain similar accuracy for an increasing number
of processes. Since incremental training and training from scratch make direct use
of data parallelism, this finding is not surprising. On the other hand, the same trend
is visible for our approach, which demonstrates that it applies global
sampling correctly at scale and therefore avoids biases. 

All approaches exhibit shorter runtimes for increasing numbers
of data-parallel processes.
Note that the gap between our approach and incremental
training does not increase with the number of processes. Instead, the gap is decreasing,
which shows that our approach is scalable and can successfully overlap the asynchronous 
updates of the rehearsal buffer and the global sampling with the training iterations, despite increasing complexity of all-to-all communication.

Note that with increasing number of processes, our approach
samples $r$ representatives and serves the same number (on the average).
Thus, the average training time of our approach is only determined by the $r$ additional representatives assembled into augmented 
mini-batches at every iteration, as shown in Section~\ref{sec:augmentation-breakdown}.

\section{Discussion}
\label{sec:discussion}

\textbf{Efficiency at Scale.}
The accumulation of representatives in the distributed rehearsal buffers
may grow to large sizes, but our approach aggregates the free memory
on the compute nodes in a scalable fashion. Specifically, given only
a fraction of the host memory on each compute node (1~GB in our experiments),
our approach was capable of storing 30\% of the ImageNet dataset even at
medium scale (128 GPUs). Furthermore, this amount of free memory can be calculated in advance as it is bounded w.r.t. the number of classes $K$ and many additional data reduction techniques can be applied if necessary
(e.g., compression). As in~\cite{verwimp2021rehearsal}, one could suspect that training repeatedly over a limited number of representatives would end up overfitting the rehearsal buffer, which may be an inherent limitation of CL. In this regard, our approach enables the aggregated size to grow proportionally with the number of processes. Thus, we retain a large and diverse set of representatives, which increases the quality of continual learning in
combination with data-parallel training beyond the limits acknowledged by
other state-of-art approaches.

\textbf{Generality.}
Our distributed rehearsal buffer stores generic tensors and supports dynamic
addition of new classes. In this paper we demonstrated its effectiveness for class-incremental classification problems.
The approach could however be easily applied to generative models (in which
case we can simply use one class to store all representatives).

\section{Conclusions}
\label{sec:conclusion}

This research contributes  to the field of CL by leveraging the concept of rehearsal buffer as a foundational element for addressing the challenges posed by evolving datasets in DL models. The concept is extended, to make it suitable for data-parallel training, thus enhancing the efficiency and scalability of DL models. We designed and implemented a distributed
rehearsal buffer that handles the selection of representative training samples, updates of the local rehearsal buffers, and the preparation of augmented mini-batches (sampled from all remote rehearsal buffers using optimized RDMA-enabled techniques) asynchronously in the background. A key aspect is the  incorporation of innovative design principles, including asynchronous techniques and the utilization of low-overhead, RDMA-aware, all-to-all communication patterns.
Extensive experiments on 128 GPUs of the ThetaGPU with 3 different models 
and a sequence of 4 tasks derived from the ImageNet-1K dataset underscore the scalability and effectiveness of our approach. As a notable result, in the best case with ResNet-50, our method
can improve the top-5 classification accuracy from 23.1\% to 80.55\% compared with incremental training, with just a small runtime increase---an ideal trade-off that combines the best of both baselines used in the comparison.

\bibliographystyle{IEEEtran}
\small{
  \bibliography{references}
}

\end{document}